\documentclass{llncs}

\usepackage{microtype}

\usepackage{booktabs} %

\usepackage{breakcites}
\usepackage{amsmath,amssymb,xcolor,graphicx,wrapfig,paralist}
\usepackage{algorithmicx,algorithm}
\usepackage[noend]{algpseudocode}
\usepackage{uppaal}
\usepackage{hyperref}
\usepackage[capitalise]{cleveref}

\usepackage{caption}
\usepackage{listings}

\usepackage{makecell}

\usepackage{xcolor}
\usepackage{color,colortbl}

\usepackage{pgfplots}
\usepackage{pgfplotstable}
\usetikzlibrary{fadings,shapes.arrows,shadows}
\tikzfading[name=arrowfading, top color=transparent!0, bottom color=transparent!95]
\tikzset{arrowfill/.style={#1,general shadow={fill=black, shadow yshift=-0.8ex, path fading=arrowfading}}}
\tikzset{arrowstyle/.style n args={3}{draw=#2,arrowfill={#3}, single arrow,minimum height=#1, single arrow,
single arrow head extend=.3cm,}}

\usepackage{subfig}

\usepackage{wrapfig}

\usepackage{ellipsis, mparhack, ragged2e} %
\usepackage[l2tabu, orthodox]{nag} %

\usetikzlibrary{arrows.meta, shapes.misc, shapes.multipart}

\usepackage{tabu,multirow}
\newcolumntype{L}{X}
\newcolumntype{R}{>{\raggedleft\arraybackslash}X}
\newcolumntype{C}{>{\centering\arraybackslash}X}

\usepackage{xparse}
\usepackage{mathtools}
\usepackage{environ}

\newcommand{\ego}{\emph{Ego}\xspace}
\newcommand{\front}{\emph{Front}\xspace}

\RequirePackage{marvosym}

\DeclarePairedDelimiter{\delimabs}{\lvert}{\rvert}
\DeclarePairedDelimiter{\delimnorm}{\lVert}{\rVert}
\DeclarePairedDelimiter{\delimpospart}{\lgroup}{\rgroup^+}

\DeclarePairedDelimiterX{\deliminner}[2]{\lange}{\rangle}{#1, #2}
\DeclarePairedDelimiter{\delimcardinality}{\lvert}{\rvert}
\DeclarePairedDelimiter{\delimset}{\lbrace}{\rbrace}
\DeclarePairedDelimiter{\delimtuple}{(}{)}
\DeclarePairedDelimiter{\delimlistt}{[}{]}
\DeclarePairedDelimiter{\delimfun}{(}{)}

\NewDocumentCommand{\abs}{sm}{\IfBooleanTF{#1}{\delimabs{#2}}{\delimabs*{#2}}}
\NewDocumentCommand{\norm}{sm}{\IfBooleanTF{#1}{\delimnorm{#2}}{\delimnorm*{#2}}}
\NewDocumentCommand{\pospart}{sm}{\IfBooleanTF{#1}{\delimpospart{#2}}{\delimpospart*{#2}}}
\NewDocumentCommand{\negpart}{sm}{\IfBooleanTF{#1}{\delimnetpart{#2}}{\delimnetpart*{#2}}}
\NewDocumentCommand{\inner}{sm}{\IfBooleanTF{#1}{\deliminner{#2}}{\deliminner*{#2}}}
\NewDocumentCommand{\cardinality}{sm}{\IfBooleanTF{#1}{\delimcardinality{#2}}{\delimcardinality*{#2}}}
\NewDocumentCommand{\set}{sm}{\IfBooleanTF{#1}{\delimset{#2}}{\delimset*{#2}}}
\NewDocumentCommand{\tuple}{sm}{\IfBooleanTF{#1}{\delimtuple{#2}}{\delimtuple*{#2}}}
\NewDocumentCommand{\closure}{sm}{\IfBooleanTF{#1}{\delimclosure{#2}}{\delimclosure*{#2}}}
\NewDocumentCommand{\listt}{sm}{\IfBooleanTF{#1}{\delimlistt{#2}}{\delimlistt*{#2}}}
\NewDocumentCommand{\fun}{smm}{\IfBooleanTF{#1}{{#2}\delimfun{#3}}{{#2}\delimfun*{#3}}}
\NewDocumentCommand{\funMacro}{smm}{\IfNoValueTF{#3}{#1}{\fun{#2}{#3}}}

\DeclareMathOperator{\ExistsOp}{\exists}
\DeclareMathOperator{\ForallOp}{\forall}

\NewDocumentCommand{\Exists}{gg}{\IfNoValueTF{#1}{\ExistsOp}{\ExistsOp #1. \, #2}}
\NewDocumentCommand{\Forall}{gg}{\IfNoValueTF{#1}{\ForallOp}{\ForallOp #1. \, #2}}

\NewDocumentCommand{\convto}{G{}}{\xrightarrow{#1}}
\NewDocumentCommand{\weakto}{G{}}{\xrightharpoonup{#1}}
\NewDocumentCommand{\weakstarto}{G{}}{\xrightharpoonup[*]{#1}}

\NewDocumentCommand{\distributions}{d()}{\funMacro{\mathcal{D}}{#1}}

\NewDocumentCommand{\actions}{d()}{{\IfNoValueTF{#1}{\mathit{Act}}{\fun{\mathit{Act}}{#1}}}}

\newcommand{\pr}{\mathbb P}
\renewcommand{\Pr}[3]{\pr^{#1}\hspace{-0.16em}\left[{#3}\right]}   %

\newcommand{\E}{\mathbb{E}}
\newcommand{\C}{{C}}
\newcommand{\U}{{U}}
\newcommand{\F}{{F}}
\newcommand{\G}{\mathcal{M}}%
\newcommand{\Rpos}{\mathbb{R}_{\geq 0}}
\newcommand{\R}{\mathbb{R}}

\newcommand{\N}{\mathbb{N}}
\newcommand{\Conf}{\mathbb{C}}

\newcommand{\juststratego}{{\sc Stratego}}
\newcommand{\conarrow}[1]{\stackrel{#1}{\rightarrow}}

\advance\textwidth2mm %
\advance\hoffset-3mm
\advance\voffset-3mm

\let\llncssubparagraph\subparagraph
\let\subparagraph\paragraph
\usepackage[compact]{titlesec}
\let\subparagraph\llncssubparagraph

\titlespacing\section{0pt}{10pt plus 4pt minus 2pt}{4pt plus 2pt minus 2pt}
\titlespacing\subsection{0pt}{12pt plus 4pt minus 2pt}{2pt plus 2pt minus 2pt}
 \newcommand{\myspace}{\vspace{-1.7em}}

\newcommand{\safestrat}{\sigma_{\mathit{safe}}}
\newcommand{\faststrat}{\sigma_{\mathit{opt}}}
\newcommand{\optstrat}{\sigma_{\mathit{opt}}}

\newcommand{\topt}{\mathcal T_{\mathit{opt}}}
\newcommand{\toptkp}{\mathcal T_{\mathit{opt}}^{k,p}}

\newcommand{\rg}{\G}%
\newcommand{\g}{\mathcal{TG}}%
\newcommand{\features}{D}

\newcommand{\classes}{A}
\newcommand{\CA}{C}

\newcommand{\vv}{\vec x}

\definecolor{color1}{HTML}{de2d26} %
\definecolor{graph1}{HTML}{e66101}
\definecolor{graph2}{HTML}{fdb863}
\definecolor{graph3}{HTML}{b2abd2}
\definecolor{graph4}{HTML}{5e3c99}

\begin{document}
\title{SOS: Safe, Optimal and Small Strategies for Hybrid Markov Decision Processes} 

\author{Pranav Ashok\inst{1}\and Jan Kretinsky\inst{1} \and Kim Guldstrand Larsen\inst{2} \and Adrien Le Co\"ent\inst{2} \and Jakob Haahr Taankvist\inst{2} \and Maximilian Weininger\inst{1}}
\institute{Technical University of Munich, Germany
  \and
  Aalborg University, Denmark}

\newcommand{\shortauthors}{Ashok et al.}%

\maketitle

\myspace

\begin{abstract}
For hybrid Markov decision processes, \uppaal\ Stratego can compute strategies that are safe for a given safety property and (in the limit) optimal for a given cost function.
Unfortunately, these strategies cannot be exported easily since they are computed as a very long list. 
In this paper, we demonstrate methods to learn compact representations of the strategies in the form of decision trees. 
These decision trees are much smaller, more understandable, and can easily be exported as code that can be loaded into embedded systems.
Despite the size compression and actual differences to the original strategy, we provide guarantees on both safety and optimality of the decision-tree strategy.
On the top, we show how to obtain yet smaller representations, which are still guaranteed safe, but achieve a desired trade-off between size and optimality.
\end{abstract} 

\section{Introduction}
Cyber-physical systems often are safety-critical and hence strong guarantees on their safety are paramount.
Furthermore, resource efficiency and the quality of the delivered service are strong requirements; 
the behaviour needs to be optimized with respect to these objectives, while of course staying within the bounds of what is still safe.
In order to achieve this, controllers of such systems can be either implemented manually or automatically synthesized.
In the former case, due to the complexity of the system, coming up with a controller that is safe is difficult, even more so with the additional optimization requirement.
In the latter case, the synthesis may succeed with significantly less effort, though the requirement on both  safety and optimality is still a challenge for current synthesis methods. 
However, due to the size of the systems, the produced controllers may be very complex, hard to understand, implement, modify, or even just output.
Indeed, even for moderately sized systems, we can easily end up with gigabytes-long descriptions of their controllers (in the algorithmic context called strategies).

In this paper, we show how to provide a more compact representation, which can yield acceptably short and simple code for resource-limited embedded devices, and consequently can be more easily understood, maintained, modified, debugged, and the requirements are better traceable in the final controller.
To this end, as the formalism for the compact representation we choose decision trees \cite{Mitchell1997}. 
This representation is typically several orders of magnitude smaller than the classical explicit description and also is known for its interpretability and understandability \cite{Mitchell1997,DBLP:journals/aai/RiddleSE94,BrazdilCCFK15}.
The resulting encoded strategy may differ from the original one, but despite that and despite being smaller, it is still guaranteed to be safe and as nearly-optimal as the original one. 
Moreover, we can trade off additional decrease in size for decrease in performance (getting farther from optimum)  to a desired degree, while maintaining safety.

\begin{figure}[t]
	\includegraphics[width=\linewidth]{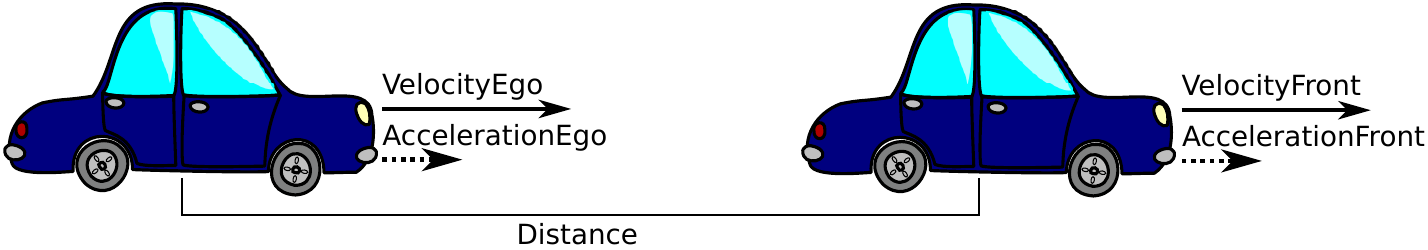}
	\caption{The two cars, \ego and \front. We control \ego and the environment controls \front. Both cars have an acceleration and a velocity. In addition, we know the distance between the cars.}
	\label{cars}
\end{figure}

\begin{example}\label{adaptive-cruise-control}
As a running example and one of the case studies, we use the following example introduced in \cite{Larsen2015} and expanded with stronger safety guarantees in \cite{cyphy2018}.

We consider two cars \ego and \front, depicted in Figure \ref{cars}. 
We control \ego, whereas \front is controlled by the environment. 
\ego is driving behind \front, and both cars have a discrete input (the acceleration) and a continuous state (the velocity). The goal of the adaptive cruise control in \ego is, first, to stay safe (by keeping the distance between the cars greater than a given safe distance), and second, to drive as close to \front as possible, i.e. to optimize the aggregated distance between the cars.

We use \tiga \cite{DBLP:conf/cav/BehrmannCDFLL07} to get a safe strategy for \ego as in \cite{cyphy2018}, and then \stratego \cite{DBLP:conf/tacas/DavidJLMT15} to learn a (near-)optimal strategy for a desired cost function, given the constraints from the safe strategy.
The resulting strategy is output as a list with almost 6 million configurations.
Using our new methods, we obtain a decision tree representing the strategy, that has only about 2713 nodes.
Additionally, we can trade performance to reduce the size even further, e.g. by increasing the aggregated distance reasonably we can reduce the size to 1247 nodes.
\end{example}

\paragraph*{Our contribution:} 
\begin{itemize}
	\item We design and implement a framework \juststratego$^+$ to transform safe and (near-)optimal strategies into their decision-tree representation, preserving safety and the same level of optimality, while being much smaller.
	\item We provide several transformations and ways to yet further decrease the size while preserving safety, but relaxing the optimality to a desired extent.
	\item We test our methods on three case studies, where we show size reductions of up to three orders of magnitude, and quantify the additional size-performance trade-off.
\end{itemize} 

Our techniques can be used to represent (finite-memory non-stochastic) strategies for arbitrary systems exhibiting non-determinism (e.g. Markov decision processes, timed/concurrent/stochastic games).
This paper demonstrates the technique on hybrid Markov decision processes, as that is the formalism used in \stratego.

\paragraph{Related Work:}
The problem of computing strategies for  hybrid systems has been extensively studied in the past years. 
Most approaches rely on abstraction techniques:
the continuous and infinite state space of the system is represented with a finite number of symbols, {\em e.g.} 
discrete points \cite{girard2012controller,rungger2016scots}, sets of states \cite{le2017improved}, etc. 
 However, it is still hard to deal with
 uncontrollable components, even though
 some approaches exist such as 
 robust control \cite{girard2012synthesis}, or contract-based design \cite{saoud2018composition}, but they usually consider the uncontrollable component as a  bounded perturbation and do not tackle stochastic behaviour. 
The tool PESSOA \cite{roy2011pessoa,majumdar2011robust} can synthesize controllers for cyber-physical systems represented by a set of smooth differential equations %
with a specification in a fragment of Linear Temporal Logic (LTL). Abstraction techniques are used in \cite{hahn2011game} for  synthesizing strategies for a class of hybrid systems that involve random phenomena
together with discrete and continuous behaviours.
Discrete, stochastic dynamical systems are considered in \cite{svorevnova2017temporal},
where the synthesis of strategies with respect to LTL objectives is made possible with an abstraction-refinement method. 
In \cite{fehnker2004benchmarks} a number of benchmarks for hybrid system verification has been proposed, including a room heating benchmark. 
In \cite{david2012evaluation} \textsc{Uppaal SMC} was applied to the performance evaluation of several strategies proposed in the benchmark. 
However, there was no focus on safety in this approach. 
In our work, the safety strategy synthesis  relies on a discretization of the continuous variables, leading to a decidable problem that can be handled by \uppaaltiga, but we furthermore provide safety guarantees for the original system with the use of a Timed Game
abstraction based on a guaranteed Euler scheme \cite{snr17}. 

In artificial intelligence, compact (factored) representations of Markov decision processes (MDPs) have been developed using dynamic Bayesian networks~\cite{BDG95,KK99}, probabilistic STRIPS~\cite{KHW94}, algebraic decision diagrams~\cite{HSHB99}, and also decision trees~\cite{BDG95}. 
For a detailed survey of compact representations see~\cite{BDH99}.
Formalisms used to represent MDPs can, in principle, be used to represent strategies as well. In particular, variants of decision trees are probably the most used~\cite{BDG95,CK91,KP99}.
Decision trees have been also used in connection with real-time dynamic programming and reinforcement learning~\cite{BD96,DBLP:conf/appinf/Pyeatt03}.

In the context of verification, MDPs are often represented using variants of (MT)BDDs~\cite{dAKNPS00,HKNPS03,MP04}, and strategies by BDDs~\cite{WBB+10}.
Learning a compact decision-tree representation of a strategy has been investigated in~\cite{LPRT10} for the case of body sensor networks, in \cite{BrazdilCCFK15}  for finite (discrete) MDPs, and in \cite{BrazdilCKT18} for finite games, but only with Boolean variables.
Moreover, these decision trees can only predict a single action for a state configuration whereas in this work, we allow the trees to predict more than one action for a single configuration.
In control theory, \cite{ZAPREEV20181} proves that the problem of computing size-optimal determinisiation of controllers is NP-complete and hence discuss various heuristic-based  determinisation algorithms.
None of these works consider the optimization aspect, which being a soft constraint enables the trade-offs.

Permissive strategies have been studied in e.g.~\cite{DBLP:journals/ita/BernetJW02,DBLP:conf/atva/BouyerMOU11,DBLP:conf/tacas/DragerFKPU14}. 
\section{Preliminaries}
\label{sec:preliminaries}

\subsection{Hybrid Markov Decision Processes}\label{sec:mdp}
We describe the mathematical modelling framework.
The correspondence to the \textsc{Uppaal} models is straightforward.

\noindent
\begin{definition}[HMDP]
A \emph{hybrid Markov decision process} (HMDP) $\G$ is a tuple $(\C,\U,X,\F,\delta)$ 
where:
\begin{enumerate}
\item the controller $\C$ is a finite set of (controllable)
  modes $C=\{c_1,\ldots,c_k\}$,
\item the uncontrollable environment  $\U$  is  a   finite  set   of
  (uncontrollable) modes $U=\{u_1,\dots,u_l\}$,
\item  $X=(x_1,\ldots,x_n)$ is  a  finite  tuple of  continuous
  (real-valued) variables, 
\item      for     each      $c\in      C$      and     $u\in      U$,
  $\F_{c,u}:\mathbb{R}_{>0}\times\mathbb{R}^X\rightarrow\mathbb{R}^X$
  is the flow-function that describes the evolution  of the continuous
  variables over time in the combined mode $(c,u)$, and
\item\label{def:prob} $\delta$ is 
    a  family  of probability functions  $\delta_{\gamma} : U \rightarrow [0,1]$, where $\gamma=(c,u,\vv)$ is a global configuration.
  More precisely, $\delta_{\gamma}(u')$ is  the probability that $u$
  in  the global  configuration  $\gamma=(c,u,\vv)$ will  change to  the
  uncontrollable mode $u'$.
\end{enumerate}
\end{definition}
In the following, we denote by $\Conf$ the  set of  global  configurations  $C\times U\times\R^X$ of the HMDP $\G$. The above notion of HMDP actually describes an infinite-state Markov Decision Process \cite{Puterman}, where choices of  mode for the controller is made periodically and choice of mode for the uncontrollable environment is made probabilistically according to $\delta$.  Note that abstracting $\delta_\gamma$ to the support $\hat{\delta_\gamma}=\{u\,|\,\delta_\gamma(u)>0\}$, turns $\G$ into a (traditional) hybrid two-player game. 
The  inclusion of  $\delta$ allows for a probabilistic refinement of the uncontrolled environment in this game.
Such a refinement is irrelevant for the purposes of guaranteeing safety; however, it will be useful for optimizing the cost of operating the system. Indeed, rather than optimizing only the worst-case performance, we wish to optimize the overall expected behaviour. 

\subsubsection*{Strategies}
  A -- memoryless and possibly non-deterministic -- strategy $\sigma$ for  the controller  $\C$ is  a function  $\sigma:\Conf\rightarrow 2^C$,
i.e.     given    the    current    configuration    $\gamma=(c,u,\vv)$, the expression $\sigma(\gamma)$ returns the set of allowed \emph{actions} in that configuration; in our setting, the actions are the controllable modes to be used for the duration of the next period.
Non-deterministic strategies are also called permissive since they permit many actions instead of prescribing one.

The evolution of system over time is defined as follows.
Let $\gamma=(c,u,\vv)$ and $\gamma'=(c',u',\vv')$.  We write $\gamma\conarrow{\tau}\gamma'$ in  case $c'=c, u'=u$ and $\vv'=\F_{(c,u)}(\tau,\vv)$.

A \emph{run} is an  interleaved  sequence  $\pi\in\Conf\times(\mathbb R \times\Conf\times\Conf\times\Conf)^*$  of  configurations  and  relative time-delays of some given period $P$: 
\[
\pi=\gamma_o\,::\,P\,::\,\alpha_1\,::\,\beta_1\,::\,\gamma_1\,::\,P\,::\,\alpha_2\,::\,\beta_2\,::\,\gamma_2\,::\,P\,::\cdots\]
Then $\pi$ is a \emph{run according to the strategy} $\sigma$ if after each period $P$ the following sequence of discrete (instantaneous) changes are made:
\begin{enumerate}
\item the value of the continuous variables are updated according to the flow of the current mode, i.e. $\gamma_{i-1}=(c_{i-1},u_{i-1},\vv_{i-1})\conarrow{P}(c_{i-1},u_{i-1},\vv_{i})=:\alpha_{i}$;
\item the environment changes to any possible new mode, i.e. $\beta_i=(c_{i-1},u_{i},\vv_{i})$ where $\delta_{\alpha_{i}}(u_{i})>0$;
\item the controller changes mode according to the strategy $\sigma$, i.e. $\gamma_i=(c_{i},u_{i},\vv_{i})$ with $c_{i}\in\sigma(\beta_{i})$. 
\end{enumerate}

\subsubsection*{Safety}

A strategy $\sigma$ is said to be \emph{safe} with respect to a set of configuration $S\subseteq\Conf$, if for any run $\pi$ according to $\sigma$ all configurations encountered are within $S$, i.e. $\alpha_i, \beta_i, \gamma_i\in S$ for all $i$ and also $\gamma'_i\in S$ whenever $\gamma_i\conarrow{\tau}\gamma'_i$ with $\tau\leq P$.
Note that the notion of safety does not depend on the actual $\delta$, only on its supports.
Recall that almost-sure safety, i.e. with probability 1, coincides with sure safety.

We use a guaranteed set-based \emph{Euler method} introduced in \cite{cyphy2018} to ensure safety of a strategy not only at the configurations where we make decisions, but also in the continuum in between them. 
We refer the reader to Appendix \ref{sec:euler} for a brief reminder of this method.

\subsubsection*{Optimality}
Under  a given  \emph{deterministic} (i.e.\ permitting one action in each configuration) strategy $\sigma$  the game  $\G$ becomes  a completely  stochastic process  $\G\upharpoonright\sigma$, inducing  a probability measure  on sets of  runs. 
In case $\sigma$ is \emph{non-deterministic} or \emph{permissive}, the non-determinism in $\G\upharpoonright\sigma$ is resolved uniformly at random.
On such a process, we can evaluate a given optimization function.
Let $H\in\N$ be a given time-horizon, and $D$ a random variable on runs, then  $\E^{\G,\gamma}_{\sigma,H}(D)\in\Rpos$ is the expected value of $D$ on the space of runs of $\G\upharpoonright\sigma$ of length\footnote{Note that there is a bijection between length of the run and time, as the time between each step, $P$, is constant.} $H$  starting in  the configuration  $\gamma$. 
As an example of $D$, consider the integrated deviation of a continuous variable, e.g. distance between \ego and \front, with respect to a given target value.

Consequently, given a (memoryless non-deterministic) safety strategy $\safestrat$ with respect to a given safety set $S$, we want to find a deterministic sub-strategy%
\footnote{i.e. a strategy that for every configuration returns a (non-strict) subset of the actions allowed by the safe strategy} $\optstrat$  that optimizes (minimizes or maximizes) 
$\E^{\G,\gamma}_{\safestrat,H}(D)$.

\subsection{Decision Trees} \label{sec:decision-trees}

From the perspective of machine learning, \textit{decision trees} (DT) \cite{Mitchell1997} are a classification tool assigning classes to data points. 
A data point is a
$d$-dimensional vector $\vec{v} = (v_1, v_2, \dots, v_d)$ of features with each $v_i$
drawing its value from some set $D_i$. If $D_i$ is an ordered set, then the
feature corresponding to it is called \textit{ordered} or \textit{numerical}
(e.g. $\mathit{velocity} \in \R$) and otherwise, it is called
\textit{categorical} (e.g. $\mathit{color} \in \{\mathit{red, green, blue}\}$). A
(multi-class) DT can represent a function $f:\prod_{i=1}^d \features_i \to \classes$ where $\classes$ is a finite set of
classes.

A (single-label) DT over the domain $D=\prod_{i=1}^d \features_i$ with labels $\classes$ is a tuple
$\mathcal{T} = (T, \rho, \theta)$, where $T$ is a finite binary tree, $\rho$
assigns to every inner node predicates of the form $x_i \sim c$ where $\sim~\in
\{\leq, =\}$, $c \in \features_i$, and $\theta$ assigns to every leaf node a list of natural numbers
$[m_1, m_2, \dots, m_{|\classes|}]$. 
For every $\vec{v}\in D$, there exists a \emph{decision path} from the root node 
to some leaf $\ell_{\vec{v}}$. 
We say that $\vec{v}$ satisfies a predicate $\rho(t)$ if $\rho(t)$ evaluates to true when its variables are evaluated as given by $\vec{v}$.
Given $\vec{v}$ and an inner node $t$ with a predicate $\rho(t)$, the decision path
selects either the \textit{left} or \textit{right} child of $t$ based on whether $\vec{v}$ satisfies $\rho(t)$ or not.
For $\vec{v}$ from the training set, we say that the leaf node
$\ell_{\vec{v}}$ \emph{contains} $\vec{v}$.
Then $m_a$ of a leaf is the number of points contained in the leaf and classified $a$ in the training set.
Further, the classes assigned by a DT to a data point $\vec{v}$ (from or outside of the training set) are given by $\arg\max\theta(\ell_{\vec{v}}) = 
\{ i \mid \forall i,j \leq \abs{\classes}.~ \theta(\ell_{\vec{v}})_i \geq \theta(\ell_{\vec{v}})_j\}$, 
i.e. the most frequent classes in the respective leaf.

Decision trees may also predict sets of classes instead of a single class. 
Such a generalization (representing functions of the type $\prod_{i=1}^d \features_i \to 2^\classes$) is called a
\textit{multi-label} decision tree. 
In these trees, $\theta$ assigns to every leaf node a list of tuples 
$[(n_1, y_1)$, $(n_2, y_2)$, $\dots$, $(n_{|\classes|}, y_{|\classes|})]$ where $n_a, y_a \in \N$ are 
the number of data points in the leaf \textit{not} labelled by class $a$ and labelled 
by class $a$, respectively. 
The (multi-label) classification of a data point is then typically given by the majority rule, i.e. it is classified as $a$ if $n_a<y_a$.

A DT may be constructed using decision-tree learning algorithms such as ID3
\cite{id3}, C4.5 \cite{c45} or CART \cite{cart}. These algorithms take as input a training set, i.e. a set of vectors whose classes are already known, and output a DT classifier. 
The tree constructions start with a single root node containing all the data points of the training set.
The learning algorithms explore all possible predicates $p = x_i \sim c$, which
split the data points of this node into two sets, $X_p$ and $X_{\neg p}$. The
predicate that minimizes the sum of entropies\footnote{Entropy of a set $X$ is $H(X) =
	\sum_{a\in A} p_a \log_2(p_a) + (1 - p_a) \log_2(1 - p_a)$, where $p_a$ is the fraction of samples in
	$X$ belonging to class $a$. See \cite{multilabeldt} for more details.} of the two sets is selected. These sets are added as
child nodes to the node being split and the whole process is repeated, splitting
each node further and further until the entropy of the node becomes 0, i.e. all
data points belong to the same class. Such nodes are called \textit{pure} nodes. 
This construction is extended to the multi-label setting by some of the algorithms. 
A multi-label node is called \textit{pure} if there is at least one class that is 
endorsed by all data points in that node, i.e. $\exists a\in A : n_a = 0$.

If the tree is grown until all leaves have zero entropy, then the
classifier memorizes the training data exactly, leading to \textit{overfitting} \cite{Mitchell1997}. This might not be
desirable if the classifier is trained on noisy data or if it needs to predict
classes of unknown data. The learning algorithms hence provide some parameters,
known as \textit{hyperparameters}, which may be tuned to generalize the
classifier and improve its accuracy.
Overfitting is not an issue in our setup where we want to learn the strategy function (almost) precisely.
However, we can use the hyperparameters to produce even smaller representations of the function, at the ``expense'' of not being entirely precise any more.
One of the hyperparameters of interest in
this paper is the minimum split size $k$. It can be used to stop splitting nodes
once the number of data points in them become smaller than $k$. By setting
larger $k$, the size of the tree decreases, usually at the expense of increasing
the entropy of the leaves. There also exist several pruning techniques
\cite{pruning-Mingers, pruning-Esposito}, which remove either leaves or entire
subtrees after the construction of the DT.

\subsection{Standard \stratego Workflow}\label{sec:stratego-workflow}

The process of obtaining an optimized safe strategy $\sigma_{opt}$ using \stratego\ is depicted as the grey boxes in Fig. \ref{fig:uppaal-stratego+-workflow}. 
First, the HMDP $\rg$ is abstracted into a 2-player (non-stochastic) timed game $\g$, ignoring any stochasticity of the behaviour.
Next, \tiga\ is used to synthesize a safe strategy $\safestrat: \Conf \to 2^\C$ for $\g$ and the safety specification $\varphi$, which is specified using a simplified version of timed computation tree logic (TCTL)\cite{DBLP:conf/cav/BehrmannCDFLL07}.
After that, the safe strategy is applied on $\rg$ to obtain $\rg\upharpoonright\safestrat$. 
It is now possible to perform reinforcement learning
on $\rg\upharpoonright\safestrat$ in order to learn a sub-strategy $\faststrat$
that will optimize a given quantitative cost, given as any run-based expression containing e.g. discrete variables, locations, clocks, hybrid variables.
For more details, see \cite{pre-stratego-learning,stratego-tool-paper}.
\section{Stratego$^+$}

In this section, we discuss the new \strategoplus framework following with each of its components are elucidated.

\subsection{New Workflow} \label{sec:stratego-plus-workflow}

\begin{figure}
	\pgfdeclarelayer{bg}    %
	\pgfsetlayers{bg,main}  %
    
	\begin{tikzpicture}[%
	scale=0.9, every node/.style={scale=0.9},
	x=1.75cm,
	y=1cm, 
	box/.style={
		fill,
		color=black!10,
		text=black,
		rectangle, 
		draw=black, 
		minimum width=1.2cm, 
		minimum height=1.2cm
	},
	edge/.style={
		-{Latex[length=2mm]},
	},
	highlight/.style={
		fill,
		color=orange!70,
		text=black,
		draw=black
	}
	]%
	
	\node[box] (1) at (0.5, 0) {$\G$};
    \node[box] (2) at (1.5, 0) {$\g$};
	\node[box] (3) at (3, 0) {$\safestrat$};
	\node[box] (0) at (4, 2) {$\G\upharpoonright\safestrat$};
	\node[box] (5) at (5.7, 2) {$\sigma_{opt}$};
	\node[box,highlight] (7) at (7.6, 2) {$\mathcal T_{opt}$};
    
	\node[box,highlight] (6) at (3, -2) {$\mathcal T_{\safestrat}^{k,p}$};
    \node[box,highlight] (4) at (4, -2) {$\rg\upharpoonright\mathcal T_{\safestrat}^{k,p}$};
	\node[box,highlight] (8) at (5.7, -2) {$\sigma_{opt}^{k,p}$};
	\node[box,highlight] (9) at (7.6, -2) {$\mathcal T_{opt}^{k,p}$};
	
	\draw[edge] (2) to node[above] {$\textsc {Uppaal}$} node[below] {$\textsc{Tiga}$} (3);
	\draw[edge] (1) to (0.5,2) to (0);
    \draw[edge] (1) to (0.5,-2.9) to (4,-2.9) to (4);
	\draw[edge] (1) to (2);
	\draw[edge] (3) to (0);
	\draw[edge] (6) to (4);
	\draw[edge] (4) to node[above] {\textit{Stratego}} node[below] {\textit{learning}} (8);
	\draw[edge] (0) to node[above] {\textit{Stratego}} node[below] {\textit{learning}}  (5);
	\draw[edge] (8) to node[above] {\textit{DT learning}} node[below] {\textit{exact}} (9);
	\draw[edge] (3) to node[left] {\textit{DT learning}} node[right] {\textit{k,p}}(6);
	\draw[edge] (5) to node[above] {\textit{DT learning}} node[below] {\textit{exact}}(7);
	
	\begin{pgfonlayer}{bg}
		\fill [yellow!25] (7.1,-3) rectangle (8.1,3);
		\node[text=gray] at (7.6,0.5) {\Large SOS};
	\end{pgfonlayer}
	
	\end{tikzpicture}
	\caption{\strategoplus workflow. The dark orange nodes are the additions to the original workflow, which now involve DT learning, the yellow-shaded area delimits the desired safe, optimal, and small strategy representations.}
	\label{fig:uppaal-stratego+-workflow}
\end{figure}
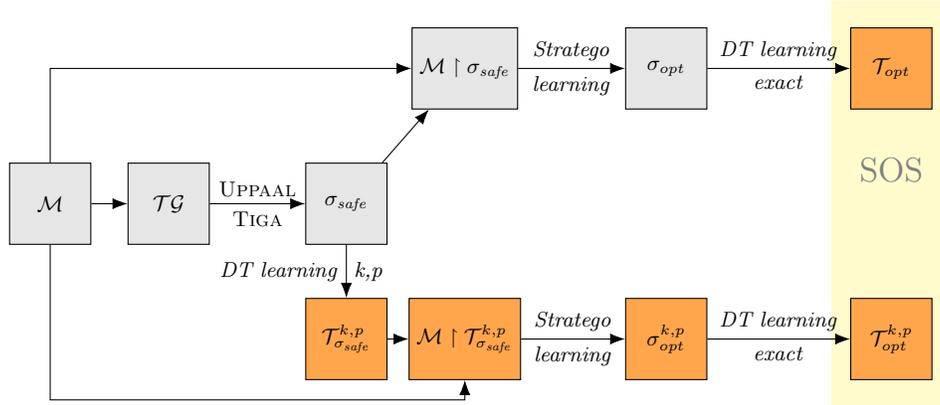

\strategoplus extends the standard workflow in two ways: 
Firstly, in the top row, we generate the DT $\mathcal T_{opt}$ that exactly represents $\optstrat$, yielding a small representation of the strategy.

The DT learning algorithm can make use of two (hyper-)parameters $k$ and $p$ which may be used to prune the DT; this approach is described in Section \ref{sec:safe}.
While pruning reduces the size of the DT, the resultant tree no longer represents the strategy exactly.
Hence it is not possible to prune a DT representing deterministic strategies, like in the case of the $\faststrat$ described in the first row of the workflow, as safety would be violated.

However, for our second extension we apply the DT learning algorithm to the non-deterministic, permissive strategy $\safestrat$, resulting in $\mathcal T_{\safestrat}^{k,p}$. 
This DT is less permissive, thereby smaller, since the pruning disallows certain actions; yet it still represents a safe strategy (details in Section \ref{sec:safe}).
Next, as in the standard workflow, this less permissive safe strategy is applied to the game and \juststratego\ is used to get a near-optimal strategy $\optstrat^{k,p}$ for the modified game $\rg\upharpoonright\mathcal T_{\safestrat}^{k,p}$.
In the end, we again construct a DT exactly representing the optimal strategy, namely $\mathcal T_{opt}^{k,p}$.
Note that in the game restricted to $T_{\safestrat}^{k,p}$ fewer actions are allowed than when it is restricted only to $\safestrat$, and hence the resulting strategy could perform worse.
For example, let $\safestrat$ allow decelerating or remaining neutral for some configuration, while $T_{\safestrat}^{k,p}$ pruned the possibility to remain neutral and only allows decelerating. 
Thus, $\optstrat$ remains neutral, whereas $\optstrat^{k,p}$ has to decelerate and thereby increase the distance that we try to minimize.

In both cases, the resulting DT is safe by construction since we allow the DT to predict only pure actions (actions allowed by all configurations in a leaf, see next section for the formal definition). We convert these trees into a nested if-statements code, which can easily be loaded onto embedded systems.

\subsection{Representing strategies using DT}

A DT with domain $\Conf$ and labels $\CA$ can learn a (non-deterministic) strategy $\sigma:\Conf \to 2^\CA$.
The strategy is provided as a
list of tuples of the form $(\gamma, \{a_1, \dots, a_k\})$, where $\gamma$ is a global
configuration and $\{a_1, \dots, a_k\}$ is the set of actions permitted by
$\sigma$.
The training data points are given by the integer configurations $\gamma\in\Conf$ (safety for non-integer points is guaranteed by the Euler method; see Section \ref{sec:mdp}) and the set of classes for each $\gamma$ is given by $\sigma(\gamma)$. 
Consequently, a multi-label decision tree learning algorithm
as described in Section \ref{sec:decision-trees} can be run on this dataset to obtain a tree $\mathcal{T}_\sigma$ representing the strategy $\sigma$.

Each node of the tree contains the set of configurations that satisfy the decision path
traced from the root of the tree to the node. 
The leaf attribute $\theta$ gives, for each action $a$, the number of configurations in the leaf where the strategy disallows and allows $a$, respectively. 
For example, consider a node with 10 configurations with $\theta = [(0, 10), (2, 8), (9, 1)]$.
This means that the first action is allowed by all 10 configurations in the node, the second
action is disallowed by 2 configurations and allowed by 8, and the third action is
disallowed by 9 configurations and allowed only by 1.

Since we want the DT to exactly represent the strategy, we need to run the learning algorithm until the entropy of all the leaves becomes 0, i.e. all configurations of the leaf agree on every action.
More formally, given a leaf $\ell$ with $n$ configurations we require $\theta(\ell) = (0,n)$ or $\theta(\ell) = (n,0)$ for every action.
We call an action that all configurations allow a \emph{pure action}.

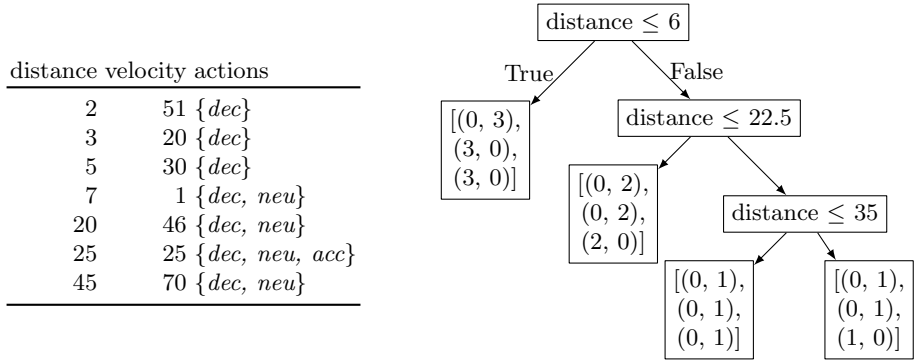
\begin{figure}[t]
	\begin{minipage}{0.4\linewidth}
		\centering
		\begin{tabular}{rr>{\em}l}
			distance & velocity & \normalfont{actions} \\
			\toprule
			2 & 51 & \{dec\} \\
			3 & 20 & \{dec\} \\
			5 & 30 & \{dec\} \\
			7 & 1 & \{dec, neu\} \\
			20 & 46 & \{dec, neu\} \\
			25 & 25 & \{dec, neu, acc\} \\
			45 & 70 & \{dec, neu\} \\
			\bottomrule
		\end{tabular}
		\label{tab:simple-dt-dataset}
	\end{minipage}\hfill
	\begin{minipage}{0.55\linewidth}
	\centering
\begin{tikzpicture}[>=latex,line join=bevel,every text node part/.style={align=center},y=0.5cm, scale=0.85]
\node (1) at (0,4) [draw=black,rectangle] {[(0, 3),\\ (3, 0),\\ (3, 0)]};
\node (0) at (2,8) [draw=black,rectangle] {distance $\leq$ 6};
\node (3) at (2,2) [draw=black,rectangle] {[(0, 2),\\ (0, 2),\\ (2, 0)]};
\node (2) at (3.5,5) [draw=black,rectangle] {distance $\leq$ 22.5};
\node (5) at (3.5,-1) [draw=black,rectangle] {[(0, 1),\\ (0, 1),\\ (0, 1)]};
\node (4) at (5,2) [draw=black,rectangle] {distance $\leq$ 35};
\node (6) at (6,-1) [draw=black,rectangle] {[(0, 1),\\ (0, 1),\\ (1, 0)]};

  \draw [->] (2) -- (3);
  \draw [->] (0) -- node[right] {False} (2);
  \definecolor{strokecol}{rgb}{0,0,0};
  \pgfsetstrokecolor{strokecol}
  \draw [->] (4) -- (6);
  \draw [->] (4) -- (5);
  \draw [->] (0) -- node[left] {True} (1);
  \draw [->] (2) -- (4);
\end{tikzpicture}

 	\end{minipage}
\caption{A sample dataset (left); and a (multi-label) decision tree generated from the dataset (right). The leaf nodes contain the list of tuples assigned by $\theta$, the inner nodes contain the predicates assigned by $\rho$}
\label{fig:simple-dt}
\end{figure}

The table on the left of Fig.~\ref{fig:simple-dt} shows a toy strategy.
Based on values of distance $d$ and velocity $v$, it  %
permits a subset of the action set $\{\mathit{dec}, \mathit{neu},
\mathit{acc}\}$.
A corresponding DT encoding is displayed on the right of Fig.~\ref{fig:simple-dt}.

\subsection{Interpreting DT as strategy}
To extract a strategy from a DT, we proceed as follows:
Given a configuration $C$, we pick the leaf $\ell_{C}$ associated with it by evaluating the predicates and following a path through the DT.
Then we compute
$\theta(\ell_{C})$ = 
$[(n_1, y_1),$ $(n_2, y_2), \dots,$ $(n_{|\classes|}, y_{|\classes|})]$ where $n_a, y_a \in \N$ 
are the number of data points in the leaf \textit{not} labelled by class $a$ and labelled 
by class $a$, respectively.
The classes assigned to $\ell_{C}$ are exactly its pure actions, i.e.
$\{a \mid (0, y_a) \in \theta(\ell_{C})\}$. 

Note that allowing only pure actions is necessary in order to preserve safety.
We do not follow the common (machine learning) way of assigning classes to the nodes based on the majority criterion, i.e. the majority of the data points in that node allow the action;
because then the decision tree might prescribe unsafe actions just because they were allowed in
most of the configurations in the node. 
This is also the reason why the DT-learning algorithm described in the previous section needs to run until the entropy of all leaves becomes 0.

\subsection{Learning smaller, yet safe DT} \label{sec:safe}

We now describe how to learn a DT for a safe strategy that is smaller than the exact representation, but still preserves safety. 
A tree obtained using off-the-shelf DT learning algorithms is unlikely to exactly represent the original strategy.\footnote{This is because DT learning algorithms are usually configured to avoid overfitting on the dataset}
We use two different methods to achieve the goal: firstly, we use the standard hyperparameter named \emph{minimum split size}, and secondly, we introduce a new post-processing algorithm called \emph{safe pruning}.
Both methods rely on the given strategy being non-deterministic/permissive, i.e. permitting several actions in a leaf.

\subsubsection*{(1) Using minimum split size} 
The splitting process can be stopped before the entropy becomes 0. 
We do this by introducing a parameter $k$, which determines the minimum 
number of data points required in a node to consider splitting it further.
During the construction of the tree, a node is usually split if its entropy is greater
than 0.
When $k$ is set to an integer greater than 2, a node is split only if both the
entropy is greater than 0 \emph{and} the number of data points (configurations) in the node is at least $k$.
The strategy given by such a tree is safe as long as it predicts only pure actions, i.e. $a$ with $n_a = 0$.
In order to obtain a fully expanded tree, $k$ may be set to 2 (in nodes with $< 2$ configurations, there is nothing to split).
For larger $k$, the number of pure actions in the leaves decreases. 
Ultimately, for too large $k$, we would obtain a tree that has some leaf nodes not containing \emph{any} pure actions. In such a case, the strategy represented by the DT would not be well-defined, as for some data point no action could be picked.
However, this can be detected immediately during the construction.

\subsubsection*{(2) Using safe pruning} \label{subsec:safe-pruning}

\begin{algorithm}[h]
	\caption{Safe Pruning}\label{alg:safe-pruning}
	\begin{algorithmic}[1]
		\Procedure{Safe-Pruning}{DT $\mathcal T_\sigma=(T,\rho,\theta)$, $p\in\mathbb N$}
		\For{$i\gets1..p$}
		\State $N \gets \{n \in T \mid \mathit{LEFT}(n) \text { and } \mathit{RIGHT}(n) \text{ are leaves}\}$ \\\Comment{Candidate nodes for pruning}
		\For{each $n \in N$}
		\State $c_\ell\gets$ $\mathit{LEFT}(n)$, $c_r\gets$ $\mathit{RIGHT}(n)$
		\If{$\theta(c_\ell)\cap \theta(c_r) \neq \emptyset$} \\\Comment{Prune and keep the common classification}
		\State Convert $n$ to a leaf node %
		\State $\theta(n)\gets\theta(c_\ell) \cap \theta(c_r)$ 
		\State Remove $c_\ell$ and $c_r$ from $T$
		\EndIf
		\EndFor		
		\EndFor
		\EndProcedure
	\end{algorithmic}
\end{algorithm}

\begin{figure}[t]
    \centering
	\tikzset{MyArrowStyle/.style={single arrow, fill=orange!70, anchor=base, align=center}}
	\begin{tikzpicture}[>=latex,line join=bevel,every text node part/.style={align=center}]
	
	\node (1) at (0,0) [draw=black,rectangle] {\textbf{A}\\(0, 7): $\mathit{dec}$\\(7, 0): $\mathit{neu}$\\ (7, 0): $\mathit{acc}$};
	\node (0) at (1,2) [draw=black,rectangle] {x $\leq$ 5};
	\node (2) at (2,0) [draw=black,rectangle] {\textbf{B}\\(0, 7): $\mathit{dec}$\\ (0, 7): $\mathit{neu}$\\ (3, 4): $\mathit{acc}$};
	
	\draw [->] (1.5,3) -- (0);
	\draw [->] (0) -- node[right] {False} (2);
	\definecolor{strokecol}{rgb}{0.0,0.0,0.0};
	\pgfsetstrokecolor{strokecol}
	\draw [->] (0) -- node[left] {True} (1);
	\node[MyArrowStyle] at (4,1) {\ \ \ };
	
	\node (C) at (6.5,1) [draw=black,rectangle] {\textbf{C}\\(0, 14): $\mathit{dec}$\\ (7, 7): $\mathit{neu}$\\ (10, 4): $\mathit{acc}$};
	\draw [->] (7,3) -- (C);

	\end{tikzpicture}
	
	\caption{Illustration of safe pruning applied to a node. The pure action of leaf \textbf{A} is just $\mathit{dec}$, for \textbf{B} it is both $\mathit{dec}$ and $\mathit{neu}$. Safe pruning replaces the nodes with \textbf{C}, where only $\mathit{dec}$ is a pure action.}
	\label{fig:safe-pruning-example}
\end{figure}
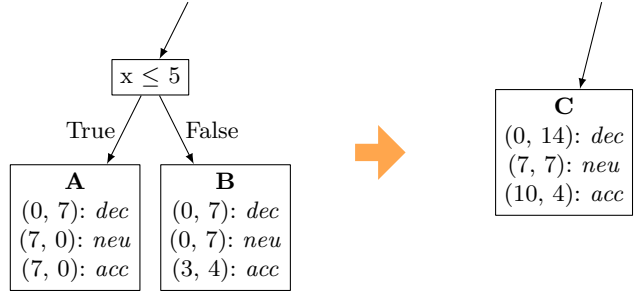

Another way of obtaining a smaller tree is by using a procedure to prune the leaves 
of the produced tree by merging them while preserving safety.
For example, consider the decision node on the left of Figure \ref{fig:safe-pruning-example} with two children that are leaves \textbf{A} and \textbf{B}. 
For \textbf{A}, only the action $\mathit{dec}$ is pure (i.e. allowed by all configurations in the leaf), while for \textbf{B} both $\mathit{dec}$ and $\mathit{neu}$ are pure. 
Since the sets of pure actions of the two leaf nodes intersect, we can safely remove both \textbf{A} and \textbf{B} and replace the decision node with a new leaf node \textbf{C} that contains only those actions that are in the intersection, in this case only $\mathit{dec}$.

Algorithm~\ref{alg:safe-pruning} describes the pruning process
formally. If $\theta$ returns only safe actions, then the tree obtained
after pruning is guaranteed to represent a safe strategy, although a
less-permissive one. The algorithm may be run for multiple (possibly 0) rounds, denoted by
$p$, at most until we get a ``fully pruned'' tree representing a safe but
deterministic strategy. 
We denote by $\mathcal{T}_{\safestrat}^{k,p}$ the decision tree for $\safestrat$ constructed by only splitting nodes with $k$ or more data points, followed by $p$ rounds of safe pruning.
Clearly, the more permissive the original strategy is, the more we can prune using safe pruning.

When generating $T_{\safestrat}^{k,p}$, we use a modified implementation of the CART decision tree learning algorithm implemented in the \texttt{DecisionTreeClassifier} class of the Python-based machine learning library Scikit-learn~\cite{scikit}. 
Since we construct the DT from a safe strategy and as long as we
let the DT-encoded strategy have at least one pure action in each leaf, the strategy
will remain safe. With this in mind, we can freely change the parameters of the 
\texttt{DecisionTreeClassifier} class. However, in our experiments, we picked  only the minimum split size $k$ from the Scikit-parameters as a demonstrative
example, as well as our newly introduced $p$.  
The methods described in this paper would work with other parameters as well. 

\subsection{Comparing DTs to Binary Decision Diagrams}\label{sec:bdd}

A Binary Decision Diagram (BDD, e.g. \cite{BDD}) is a popular data structure that can be used to
represent boolean functions $f: \mathbb{B}^n \to \mathbb{B}$. It may also be
used to represent strategies by encoding configurations and actions into a suitable form
via bit-blasting, i.e. converting them into propositional formulae. For example,
the configuration-action pair $((x=6, y=2), a_0)$ can be represented as $(x_2 \land x_1
\land \neg x_0 \land \neg y_2 \land y_1 \land \neg y_0 \land a_0)$, if it is
known that the maximum value that $x$ and $y$ can take is less than 8 (3 bits).
A strategy can be seen as a disjunction $\bigvee_{\gamma,a\in\sigma(\gamma)} (\gamma,a)$ of all configuration-action pairs $(\gamma,a)$ permitted by the strategy $\sigma$.
Such an encoding allows for an easy conversion into a BDD. Though theoretically 
straightforward, there are some practical concerns involved when constructing
the BDD. Mainly, the ordering of the variables in the BDD can drastically change
its size. While computing the optimal ordering so as to have the smallest BDD is
an NP-complete problem \cite{bddordering}, various heuristics exist that can
be used to get better orderings. We use the CUDD package \cite{somenzi2009cudd}
to construct the BDD, along with Rudell's Sifting  
reordering technique \cite{rudell-bdd-reorder}.

The main disadvantage of DTs compared to BDDs is that isomorphic subgraphs are not merged (DTs are trees, BDDs are directed acyclic graphs);
and even if merging was allowed, it would not save much.
Indeed, since DT may choose different predicates on the same level (which is an advantage in contrast to BDD with a fixed variable ordering) isomorphic subgraphs occur rarely.
There are further advantages of DT, related to learning, that make them more compact than BDD in some contexts, e.g. \cite{BrazdilCCFK15,BrazdilCKT18}.
Firstly, they can be learnt fast, using the entropy-based heuristic, compared to the graph processing and variable re-ordering of BDDs.
Secondly, a DT can ignore ``don't-care inputs''; these inputs are encodings of things that are not valid configuration-action pairs, in the sense that either the action is not available in the configuration or that it is not a valid configuration at all. 
In contrast, a BDD has to explicitly either allow or disallow these inputs.
Thirdly, DT learning can also be used to represent the strategy imprecisely using a smaller DT, which can be model checked for safety. For the modifications described in Section \ref{sec:safe}, we do not even need to re-verify safety, because this property is preserved by both our size reduction techniques.
Fourthly, DT can use much wider class of predicates, compared to single bit tests for a bit representation in a BDD.
This final point is also a reason (together with the smaller size) why DT is a more understandable representation than a BDD \cite{BrazdilCCFK15,BrazdilCKT18}. 
We also illustrate this point on a case-study in Remark \ref{rem:handcraft}.

\section{Case Studies and Experimental Results}\label{sec:experiments}

In this section, we evaluate the techniques discussed above on three different
case studies: (1) the adaptive cruise control model introduced in the motivation; (2) a two tank case study introduced in \cite{twotanks}; and (3) the heating 
system of a two room apartment adapted from~\cite{girard2013low}.

Table \ref{tab:sizes} compares representations for our case studies obtained in different ways.
We discuss results for the three case studies, denoted \texttt{cruise}, \texttt{twotanks}, and \texttt{tworooms} respectively.
Additionally, the first line displays \texttt{cruise} without the integrated Euler method, to illustrate the effect of Euler method on the final size.
All the representations are safe and as optimal as $\optstrat$ produced by \stratego.

For each of the models we display the following information:
the third column lists the number of items in the explicit list representation of $\optstrat$ output by \stratego.
The fourth column lists the number of those items that are actually relevant, i.e. sets of configurations where an actual decision is to be made.
The fifth and sixth column list the sizes of BDD and DT representations learnt from $\optstrat$, i.e.\ %
the upper path in Fig.~\ref{fig:uppaal-stratego+-workflow}.
For BDDs, since the initial ordering plays a role in the size of the final result despite applying the re-ordering heuristics, we ran 40 experiments for each model with random initial variable orderings. For creating BDDs, we used the free Python library \textit{tulip-control/dd} as an interface to CUDD.

We conclude that both BDDs and DTs reduce the size by several order of magnitude.
DTs are slightly better in all cases, and 2 orders of magnitude smaller in the tworooms model.
Note that reliably achieving good results when constructing the BDD relies on repeating the construction several times;
since already constructing a single BDD and applying the heuristics \cite{rudell-bdd-reorder} already took roughly 10 times longer than DT learning, DT can be obtained one or two orders of magnitude faster than BDDs, depending on how many times one tries constructing the BDD.
Further, for the two tanks, only DT realizes that the strategy is actually trivial.
The main reason for BDD not to spot this is the point of ignoring ``don't-care'' inputs addressed in Section \ref{sec:bdd}. 

\begin{table}[t]\centering\small\caption{Sizes of the different representations: explicit list as output by \stratego, the relevant part of the list, BDD displaying [minimum/median/maximum] over the 40 trials, and DT according to %
the upper path in Fig.~\ref{fig:uppaal-stratego+-workflow}.}
\setlength{\tabcolsep}{3pt}
\label{tab:sizes}\makebox[\linewidth]{\begin{tabular}{lrrrrr} \toprule
& \#Variables & Stratego list & List & BDD[min/med/max] & DT $\topt$ size\\ \midrule 
\texttt{cruise}$_{\texttt{non-Euler}}$%
& 5 & 1,790,034 & 308,216 & [3,718/5,066/5,890] & 2,899 \\ \midrule
\texttt{cruise}%
& 7 & 5,931,154 & 304,752 & [3,470/4,728/4,742] & 2,713 \\ \midrule
\texttt{twotanks}%
& 9 & 23,182 & 23,182 & [65/69/91] & 1 \\ \midrule
\texttt{tworooms}
& 11 & 1,924,708 & 509,715 & [16,370/20,214/25,909] & 487 \\ \bottomrule
\end{tabular}
}\end{table} %

\begin{table}[t]
	\centering\small\caption{Tables displaying the number $|\toptkp|$ of nodes of $\toptkp$ (left) and the expected performance $\E^{\G,\gamma}_{\sigma,H}(D)$ (right) for various $k$ and $p$, i.e.\ using the bottom path of Fig.~\ref{fig:uppaal-stratego+-workflow}, for the \texttt{cruise} model. Higher performance corresponds to a lower number.}\label{tab:kpE}
	\begin{minipage}{0.5\linewidth}
		\centering
		\includegraphics{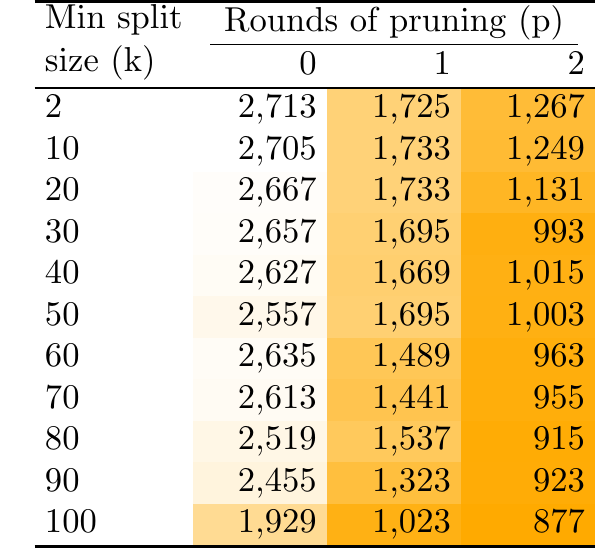}
	\end{minipage}\hfill
	\begin{minipage}{0.5\linewidth}
		\centering
		\includegraphics{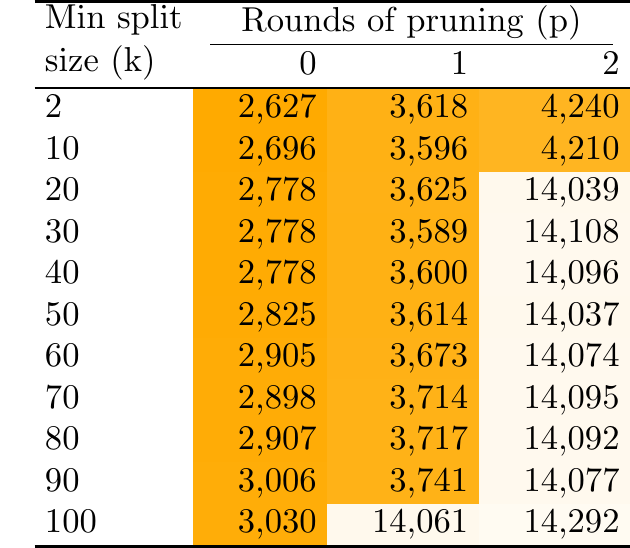}
	\end{minipage}
\end{table}

\begin{table}[t]
	\caption{Tables displaying the number $|\toptkp|$ of nodes of $\toptkp$ (left) and the expected performance $\E^{\G,\gamma}_{\sigma,H}(D)$ (right) for various $k$ and $p$, i.e.\ using the bottom path of Fig.~\ref{fig:uppaal-stratego+-workflow}, for the \texttt{tworooms} model. Higher performance corresponds to a lower number.}
	\label{tab:tworooms}
	\begin{minipage}{0.5\linewidth}
		\centering
		\includegraphics{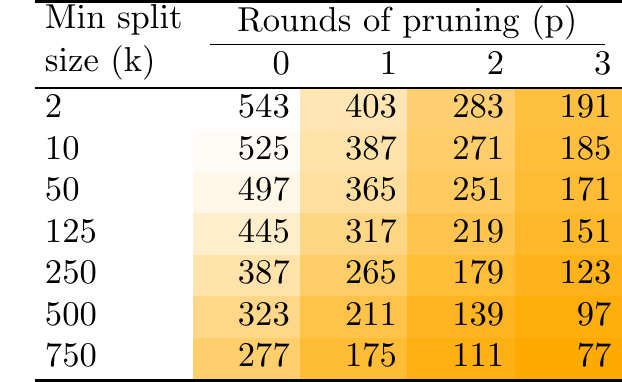}
	\end{minipage}\hfill
	\begin{minipage}{0.5\linewidth}
		\centering
		\includegraphics{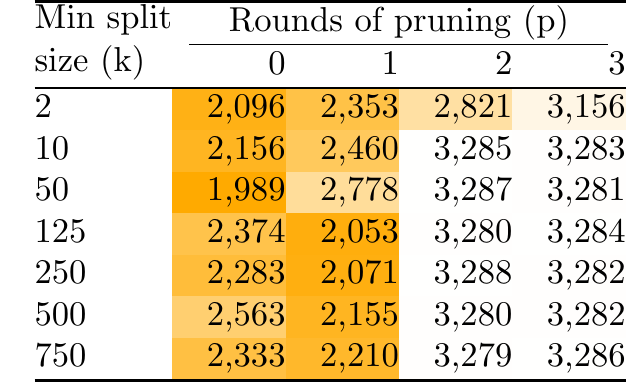}
	\end{minipage}
\end{table} 
Table \ref{tab:kpE} shows how the size of the DT can be further reduced by the bottom path of Fig.~\ref{fig:uppaal-stratego+-workflow}, when the ``exact representation'' criterion is relaxed. 
It displays the performance, i.e. the aggregated distance to \front car, and size of $\toptkp$ for different combinations of the pruning parameters $k$ and $p$.
Recall that using no pruning ($k=2, p=0$) 
yields the same DT as the upper path of Fig.~\ref{fig:uppaal-stratego+-workflow}, i.e. $\topt^{2,0}=\topt$. 

We observed that for \texttt{cruise}, increasing the values of $k$ and $p$ buys a reduction in size of the DT against a reduction in performance. For instance, using $k=80, p=0$, one can decrease the size to 2485 (by 8.4\%) while deteriorating the performance to 2907 (by 10\%). 
Allowing for half the performance (double the aggregated distance), one can make the DT even smaller than half of its original size, e.g. by setting $k=10, p=2$. 
The shading and colouring of the table display different ``trade-off zones'', each with comparable savings/losses.
The same conclusions hold for \texttt{cruise}$_\texttt{non-Euler}$, see the similar Table \ref{tab:nonEuler} in Appendix \ref{app:additional-tables}.
For \texttt{tworooms} (Table \ref{tab:tworooms}), the best performance is observed not with $k=2, p=0$, but with $k=50, p=0$. 
We conjecture that the less permissive safe strategy assists \juststratego\ in performing the optimisation faster by reducing the size of the search space.
As a result, here we get a both smaller and more performant strategy.
In the case of \texttt{twotanks}, already $\topt$ has only a single node, hence no further reductions are possible. 

\begin{remark}\label{rem:handcraft}
Interestingly, domain knowledge can reduce the DT size further and make the representation more understandable. 
Indeed, for the \texttt{cruise} model we were able to construct a DT with only 25 nodes, designing our predicates based on the car kinematics.
For example, the expected time until the front car reaches minimal velocity if it only decelerates from now on (1) plays an important role in the decision making and (2) can be easily expressed by solving the standard kinematics equation $v(t) = v_\text{current} - a_\text{dec} \cdot t$.
The resulting DT (illustrated in Appendix \ref{app:handcraft}) is thus very small and easy to interpret, as each of the few nodes has a clear kinematic interpretation.
The DT thus open the possibility for strategy representation to profit from predicate/invariant synthesis.
\end{remark} 

\section{Conclusion}

We have provided a framework for producing small representations of safe and (near-)optimal strategies, without compromising safety.
As to (near-)optimality, we can choose between two options: (i) not compromising it, or (ii) finding a suitable trade-off between compromising it (causing drops of performance) and additional size reductions.
Compared to the original sizes, we achieve orders-of-magnitude reductions, allowing for efficient usage of the strategies in e.g. embedded devices.
Compared to BDD representation, the size of the DT representation is smaller and can be computed faster; additionally trivial solutions are represented  by trivial DTs.
DTs are more readable as argued in \cite{BrazdilCKT18,BrazdilCCFK15}. 

A detailed examination of the latter point in the hybrid context remains future work.
Further, candidates for more complex predicates could be automatically generated  based on given domain knowledge or learnt from the data similarly to invariants from program runs \cite{invariants, geometric}. 
As illustrated in Remark~\ref{rem:handcraft}, this could lead to further reduction in size and improved understandability.
Additionally, isomorphic/similar subtrees could be merged as in decision diagrams and further optimizations for algebraic decision diagrams \cite{ZAPREEV20181} could be employed.
Finally, we plan to visualize the DT representation of the strategies directly in \strategoplus for convenience of the users.

\bibliographystyle{abbrv}
\bibliography{ref,related-work}
%
\appendix
\section{Appendix}

\subsection{Description of case studies}

\paragraph{Two rooms model.}This case study is based on a simple model of a two-room apartment, heated by one heater in each room. In this example, the objective is
to control the temperature of both rooms by switching on or off the heaters in the rooms. There is heat exchange
between the two rooms and with the environment (external temperature, which varies randomly within a set).

\paragraph{Cruise Control.}
Two cars {\em Ego} and {\em Front} are driving on a road as shown in Figure~\ref{cars}.
We are capable of controlling {\em Ego} but not {\em Front}. Both cars can drive a maximum of 
20 m/s forward and a maximum of 10 m/s backwards. The cars have three different 
possible accelerations: -2 m/s$^2$, 0 m/s$^2$ and 2 m/s$^2$, between which they can switch instantly. 
For the cars to be safe, there should be a distance of at least 5 m between them. Any distance less than 5 m between the cars is considered unsafe.
{\em Ego}'s sensors can detect the position of {\em Front} only within 200 meters. If the distance between the cars is more than 200 meters, then {\em Front} is considered to be {\em far away}. 

In this example, the aim is to synthesize a strategy for controllable car {\em Ego} such that it always stays far enough from uncontrollable car {\em Front}.

For the adaptive cruise control model, we differentiate between the model that is only safe at time points where an action is possible, called \texttt{cruise}$_\texttt{non-Euler}$; and the model improved with the Euler method to be safe at all time points between two actions, called just \texttt{cruise}.

\paragraph{Two Tanks.}
\begin{figure}
	\begin{center}
		\includegraphics[width=0.18\textwidth,trim={1cm 0 0 1cm},clip]{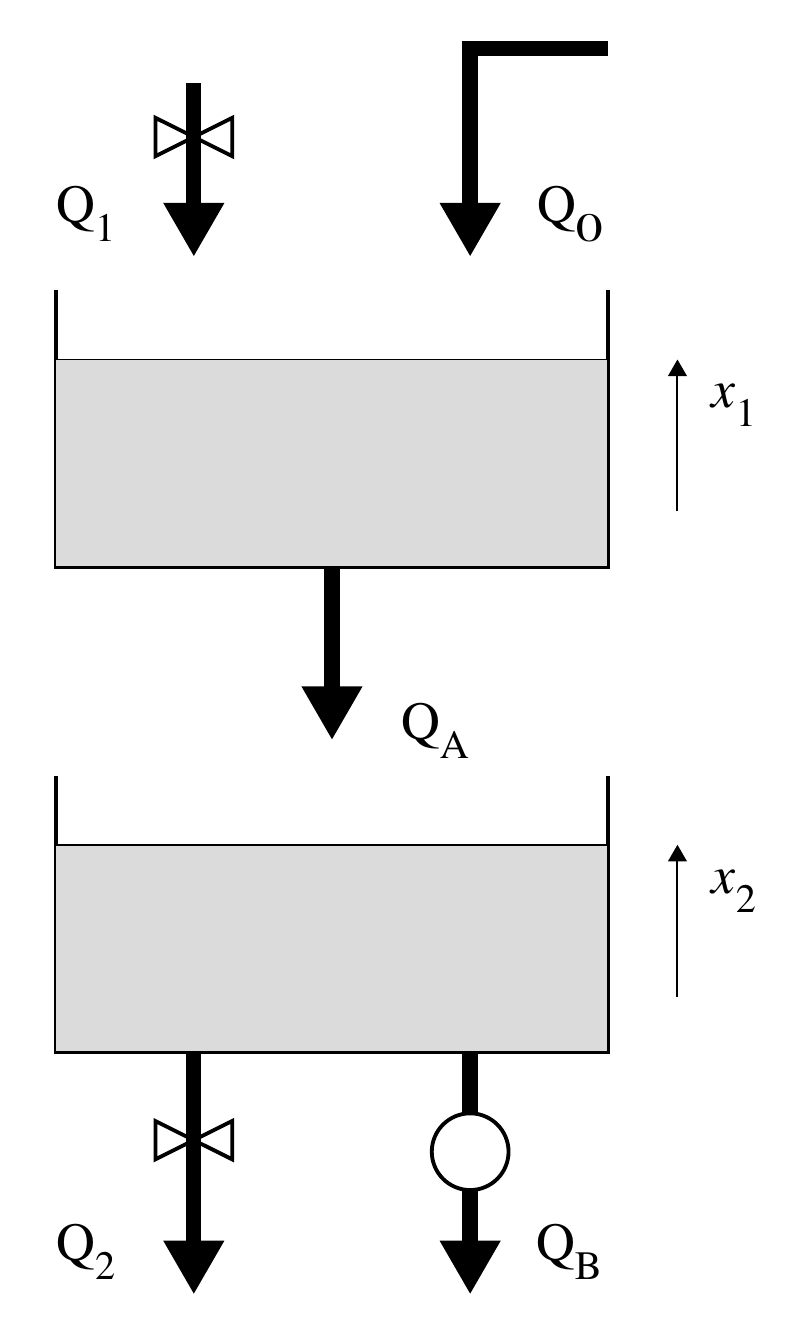}
		\caption{The two tank example.}
		\label{fig:twotanks}
	\end{center}
\end{figure}

The two-tank  system, illustrated in Figure \ref{fig:twotanks}, is a linear example taken from \cite{twotanks}. 
The system consists of two tanks and two valves.
The first valve ($Q_1$ in Fig. \ref{fig:twotanks}) adds to the inflow of tank 1 and the second valve ($Q_2$) is a drain valve for tank 2. 
There is a constant outflow from tank 2 caused by a pump ($Q_B$), as well as a constant inflow into tank 1 ($Q_0$).
There is also a flow from tank 1 to tank 2 ($Q_A$) which depends on the water level in tank 1.  The system is linearised at a desired
operating point. The objective is to keep the water level in both tanks 
within limits using a discrete open/close switching strategy for the valves. 

\subsection{Guaranteed over-approximation using the Euler Method}
\label{sec:euler}

We briefly recall the technique used in \cite{cyphy2018} to compute strategies that are safe at all time points, not just at multiples of the period $P$.

Let us consider an HMDP $\G %
= (\C,\U,X,\F,\delta)$  where the dynamics of each continuous variable $x_i \in X$ is subject to a differential equation of the form
\begin{equation}
	\dot x_i = f_i(c,u,\vec x),
    \label{eq:diff_eq}
\end{equation}
with $c \in C$ and $u \in U$. 
For every $\gamma=(c,u,\vec x)$ and $\tau\in \mathbb R$, the update $\gamma\conarrow{\tau}(c,u,\F_{(c,u)}(\tau,\vv))$ 
must match the solution of the differential equation \eqref{eq:diff_eq}, {\em i.e.}
\begin{equation}
	\F_{(c,u)}(\tau,\vv) = \vv + \int_0^\tau f(c,u,\vv) dt.
    \label{eq:integral_form}
\end{equation}
It is easy to compute the solution of a linear differential equation~\cite{butcher2016numerical}.
However, in the general case, computing the integral on the right-hand side of \eqref{eq:integral_form} is not possible exactly. 
Approximate solutions can be obtained with the use of numerical schemes, such as Euler or Runge-Kutta schemes, but in order to ensure absolute safety of the system, they should be associated with guaranteed error bounds. 
Furthermore, one has to ensure that the system is safe between time steps ($\gamma'\in S$ whenever $\gamma_i\conarrow{\tau}\gamma'$ with $\tau\leq P$).
In order to achieve that, 
\cite{cyphy2018} proceeds as follows.
First, the given HMDP $\G=(\C,\U,X,\F,\delta)$ is extended into an HMDP $\G' = (\C,\U,(X^{\min},X^{\max}),\F',\delta')$  where $X^{\min}=(x_1^{\min},\ldots,x_n^{\min})$ and
$X^{\max}=(x_1^{\max},\ldots,x_n^{\max})$ will form a lower and an upper bound on the actual valuation of $X$ at all times, 
and $\delta'$  is an abstract transition function for the opponent.
From a given global configuration $\gamma' = (c,u,(\vec x^{min}, \vec x^{max}))$ of $\G'$, $\delta'$ enables all possible transitions  that have non zero probability from $\gamma = (c,u,\vec x)$ in $\G$ for some value of $\vec x$ between $\vec x^{min}$ and $\vec x^{max}$, 
{\em i.e.} $\delta' = \{ u | \delta_{(c,u,\vec x)}(u)>0, \vec x^{min} \leq \vec x \leq \vec x^{max} \}$.
Then the flow function  $\F'_{c,u}:\mathbb{R}_{>0}\times\mathbb{R}^{(X_{\min},X_{\max})}\rightarrow\mathbb{R}^{(X_{\min}, X_{\max})}$ implements a guaranteed (set-based) Euler method which ensures the following property: whenever
\begin{itemize}
\item $(\vec w^{\min},\vec w^{\max})=\F'_{(c,u)}(P,(\vec v^{\min},\vec v^{\max}))$ and
\item $\vec w=\F_{(c,u)}(t,\vec v)$ for some $t \leq P$ and $\vec v^{\min}\leq \vec v \leq \vec v^{\max}$
\end{itemize}
then also
$ \vec w^{\min} \leq \vec w \leq \vec w^{\max}$.
For further details, we refer the reader to \cite{snr17,cyphy2018}.

\begin{figure}[htbp]
\centering
\includegraphics[width=\textwidth]{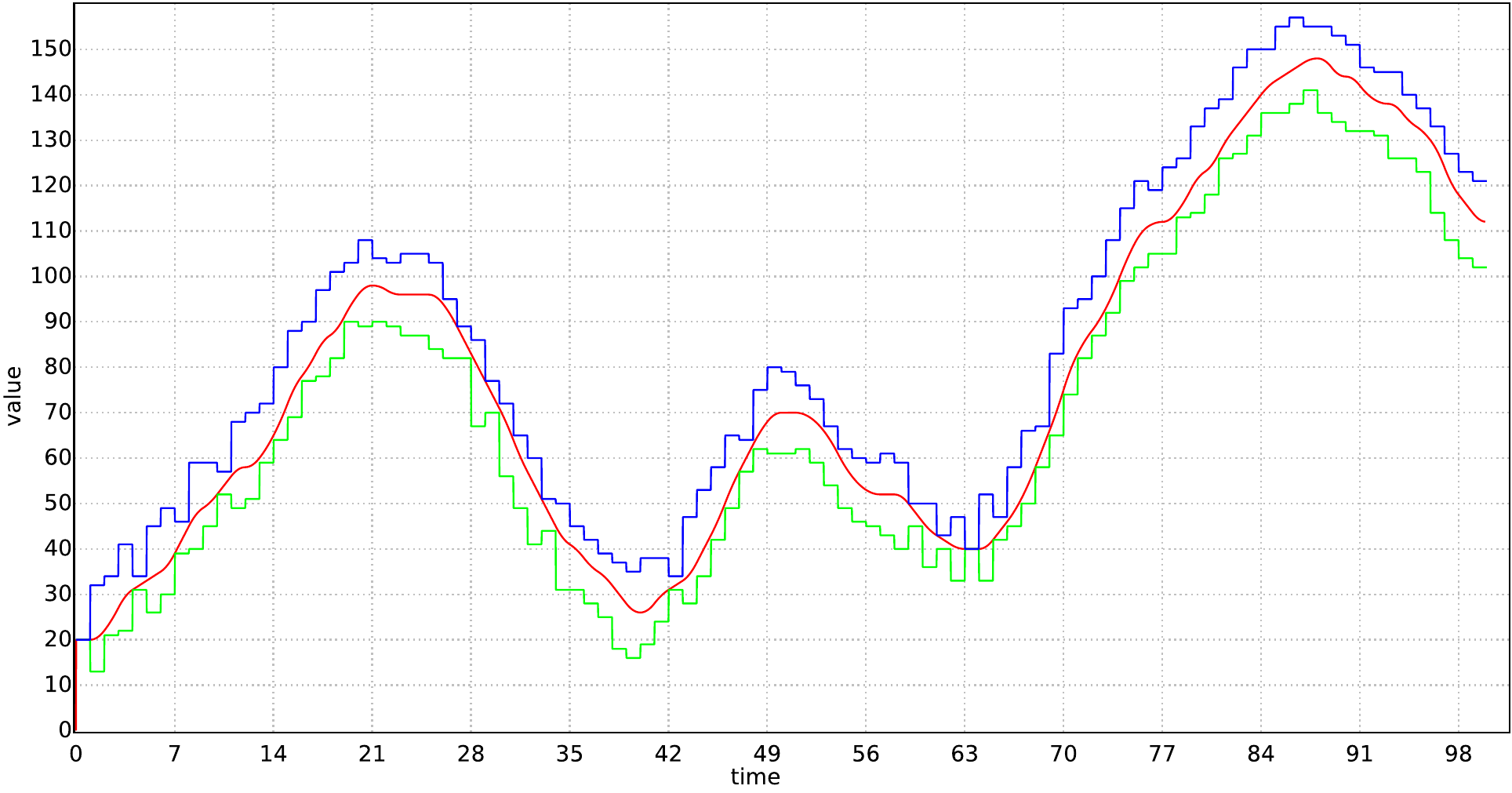}
\caption{Simulation of the distance between the cars, the red line is the continuous trajectories of the HMDP $\G$, the blue and green lines are the integer trajectory of the timed game $\mathcal {TG}$ bounding it within time.}
\label{fig:simulation_guaranteed}
\end{figure}

From the augmented HMDP $\G'$, we can build a standard timed game \cite{cyphy2018} $\mathcal{TG}$ by under-approximating $X^{\min}$ to its floor integer part, and $X^{\max}$ to its ceiling integer part.  
Consequently, one can use \tiga to synthesize a strategy that ensures the safety of the initial HMDP.
A simulation showing the bounding of the continuous trajectory of $\G$ by its timed-game abstraction $\mathcal{TG}$ is given in Fig.~\ref{fig:simulation_guaranteed}.

\clearpage

\subsection{More experimental results} \label{app:additional-tables}

\begin{figure}
	\centering
	\includegraphics[width=\textwidth]{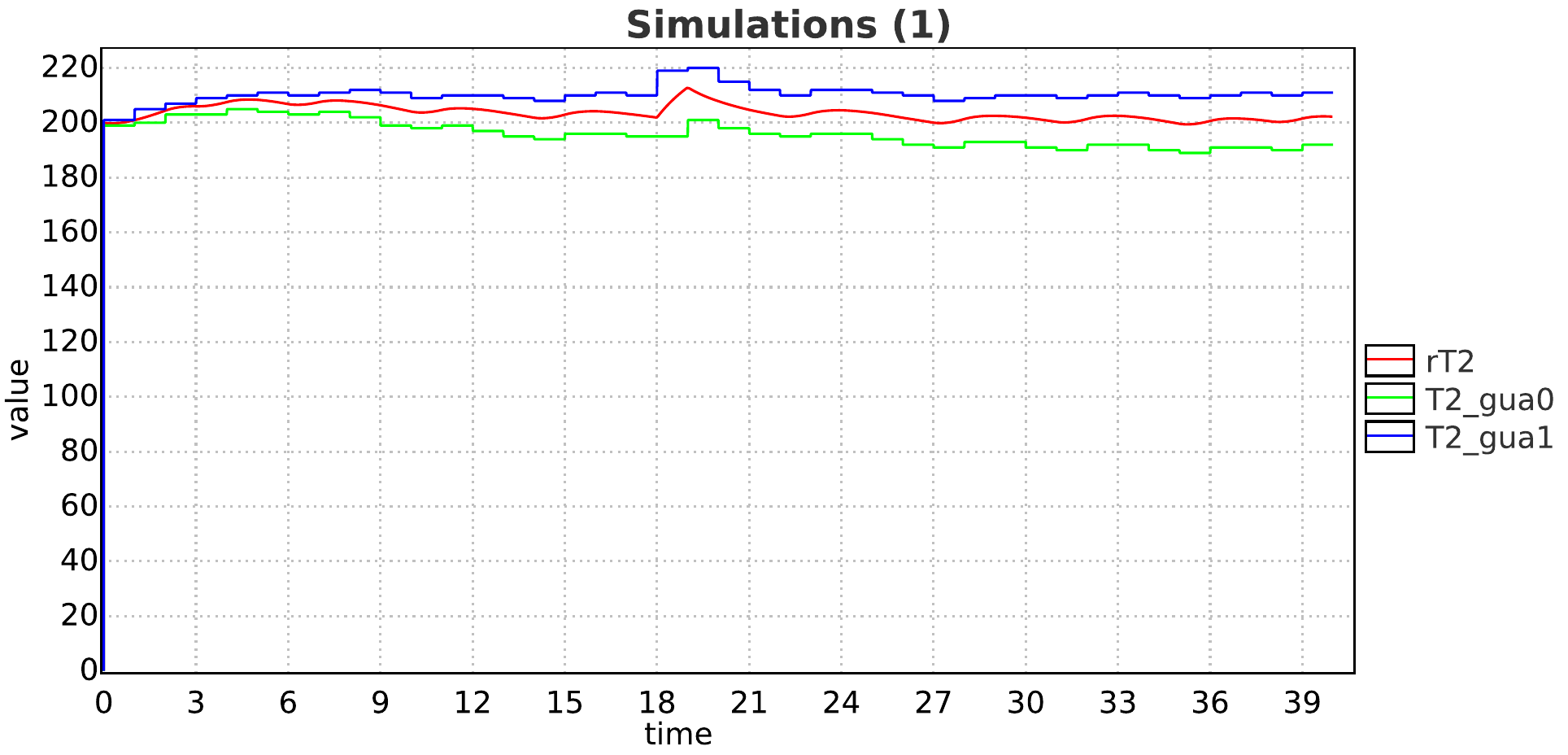}
	\caption{Simulation of the temperature of Room 2 \texttt{rT2} (red line). The blue and green lines are the integer trajectory of the timed game $\mathcal {TG}$ bounding it within time.}\label{fig:temp}
\end{figure}

Table \ref{tab:nonEuler} shows the trade-off between size and performance for the \texttt{cruise}$_\texttt{non-Euler}$ model. Figure \ref{fig:temp} shows a simulation run of \texttt{tworooms}.

\begin{table}[htbp]
	\caption{Tables displaying the number $|\toptkp|$ of nodes of $\toptkp$ (left) and the expected performance $\E^{\G,\gamma}_{\sigma,H}(D)$ (right) for various $k$ and $p$, i.e.\ using the bottom path of Fig.~\ref{fig:uppaal-stratego+-workflow}, for the \texttt{cruise}$_\texttt{non-Euler}$ model. A lower corresponds to a higher performance.}\label{tab:nonEuler}
	\begin{minipage}{0.5\linewidth}
		\centering
		\includegraphics{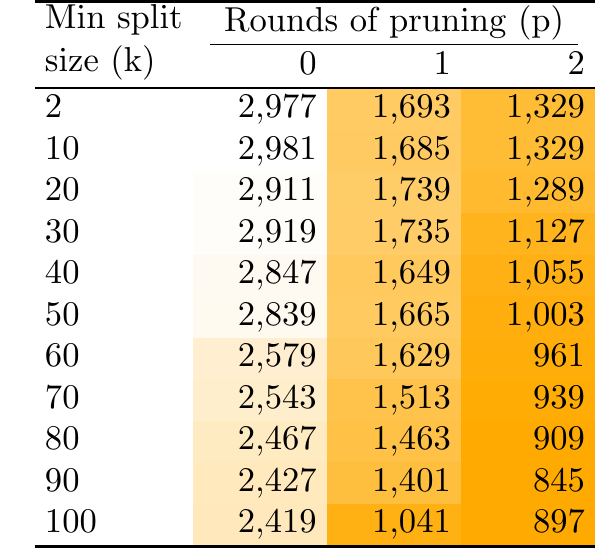}
	\end{minipage}\hfill
	\begin{minipage}{0.5\linewidth}
		\centering
		\includegraphics{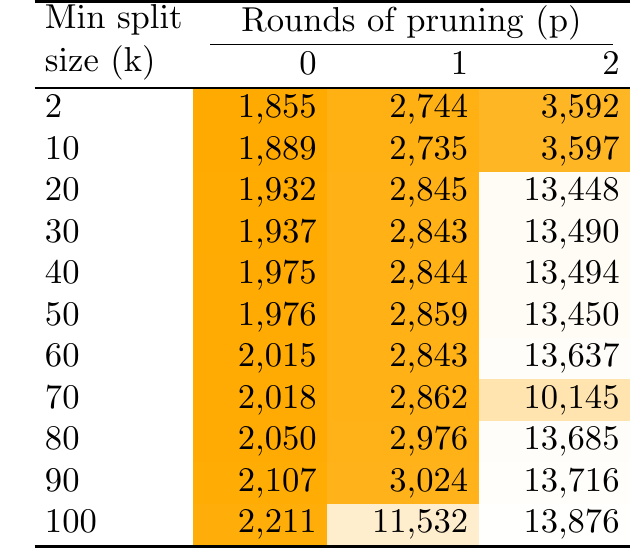}
	\end{minipage}
\end{table}

\clearpage

\subsection{Handcrafted Small DT for Adaptive Cruise Control}\label{app:handcraft}

\lstdefinestyle{customc}{
	belowcaptionskip=1\baselineskip,
	breaklines=false,
	language=C,
	showstringspaces=false,
	basicstyle=\footnotesize\ttfamily,
	keywordstyle=\bfseries\color{green!40!black},
	commentstyle=\itshape\color{purple!40!black},
	identifierstyle=\color{blue},
	stringstyle=\color{orange},
}

\lstset{escapechar=@,style=customc}

\begin{lstlisting}[caption=Handcrafted C-code describing the safe and optimal strategy for \texttt{cruise}, style=customc]
bool strategy(int act, int vEgo, int d, int vFront, int aFront, int aEgo) {
	if (check(d, vEgo, vFront, 2, aFront) > 5) {
		return act == 1; // accelerate
	} else if (check(d, vEgo, vFront, 0, aFront) > 5) {
		return act == 0; // neutral
	} else {  
		assert (check(d, vEgo, vFront, -2, aFront) > 5);
		return act == 2; // decelerate
	}
}

float check(int d, int vEgo, int vFront, int aEgo, int aFront) {
	// One unit of aEgo while front plays worst case
	float d1, d2, d3, t1, t2, vFrontNext, vEgoNext;

	t1 = (vFront + 10)/2;
		
	if (t1 > 0.5) {
		d1 = d_t(d, vEgo, vFront, aEgo, -2, 1);
		vFrontNext = vFront - 2;
		vEgoNext = vEgo + aEgo;
	} else {
		d1 = d_t(d, vEgo, vFront, aEgo, 0, 1);
		vFrontNext = vFront;
		vEgoNext = vEgo + aEgo;
	}
	
	// Decelerating while front decelerates to -10
	if (t1 > 1) {
		d2 = d_t(d1, vEgoNext, vFrontNext, -2, -2, t1 - 1);
		vEgoNext = vEgoNext - 2*(t1 - 1);
	} else {
		d2 = d1;
	}

	// Decelerating while front stays at -10
	t2 = (vEgoNext + 10)/2;
	if (t2 > 0) {
		d3 = d_t(d2, vEgoNext, -10, -2, 0, t2);
	} else {
		d3 = d2;
	}

	return d3;
}

float d_t(int d, int vEgo, int vFront, int aEgo, int aFront, int t) {
	return d + 0.5*(aFront - aEgo )*t*t 
			 + (vFront - vEgo )*t;
}
\end{lstlisting} 
\end{document}